\documentclass[letterpaper,aps,prd,twocolumn,tightenlines,preprintnumbers,nofootinbib,showkeys,superscriptaddress]{revtex4-1}
\pdfoutput=1

\usepackage{XCharter}
\usepackage{mathptmx}
\usepackage[T1]{fontenc}

\usepackage{fullpage}
\usepackage{amsfonts}
\usepackage{amsmath}
\usepackage{slashed}
\usepackage{amssymb}
\usepackage{graphicx}
\usepackage{epic}
\usepackage{eepic}
\usepackage{epsfig}
\usepackage{latexsym}
\usepackage[dvipsnames]{xcolor}
\usepackage[export]{adjustbox}
\usepackage{float}
\usepackage{multirow}
\usepackage{hyperref}
\usepackage{enumitem}
\usepackage{lipsum}
\hypersetup{colorlinks=true,citecolor=red,linkcolor=NavyBlue,urlcolor=NavyBlue}
\usepackage[caption=false,labelformat=simple]{subfig}
 
\usepackage{natbib}
\usepackage{relsize}
\usepackage[left=1.65cm,right=1.65cm,top=1.5cm,bottom=1.6cm]{geometry}
\usepackage{comment}
\usepackage{shorthand}
\usepackage{dcolumn}
\usepackage{tikz} 
\usetikzlibrary{shapes,arrows,positioning,automata,backgrounds,calc,er,patterns}
\usepackage[compat=1.1.0]{tikz-feynman}

\begin{document}
\relscale{1.05}

\title{Combined explanation of $W$-mass, muon $g-2$, $R_{K^{(*)}}$ and $R_{D^{(*)}}$  anomalies in a singlet-triplet scalar leptoquark model}

\author{Arvind Bhaskar}
\email{arvind.bhaskar@research.iiit.ac.in}
\affiliation{Center for Computational Natural Sciences and Bioinformatics, International Institute of Information Technology, Hyderabad 500 032, India}

\author{Anirudhan A. Madathil}
\email{anirudhan17@iisertvm.ac.in}
\affiliation{Indian Institute of Science Education and Research Thiruvananthapuram, Vithura, Kerala, 695 551, India}

\author{Tanumoy Mandal}
\email{tanumoy@iisertvm.ac.in}
\affiliation{Indian Institute of Science Education and Research Thiruvananthapuram, Vithura, Kerala, 695 551, India}

\author{Subhadip Mitra}
\email{subhadip.mitra@iiit.ac.in}
\affiliation{Center for Computational Natural Sciences and Bioinformatics, International Institute of Information Technology, Hyderabad 500 032, India}

\begin{abstract}\noindent
In addition to the long-standing anomalies seen in the muon $g-2$, $R_{K^{(*)}}$, and $R_{D^{(*)}}$ observables by various independent experiments, the CDF Collaboration has found another significant one in the $W$-boson mass. These anomalies might be intertwined at a fundamental level and have a single new-physics explanation. In this paper, we present a simultaneous solution to these anomalies with two scalar leptoquarks of roughly equal mass---one weak-singlet $S_1$ and the other a weak-triplet $S_3$---which mix through a Higgs portal. The solution has only a few new couplings and TeV-scale leptoquarks, making the solution testable at the LHC. We put a lower and upper bound on the leptoquark masses in this setup from the current LHC data and the assumption that all new couplings are within the perturbative limit.\\
\end{abstract}

\maketitle 
\medskip

\section{Introduction}\label{sec:intro}
\noindent
Even though the direct experimental searches for TeV-scale new physics have been unsuccessful, over the past few years, various anomalies reported by different experimental collaborations have created excitement in the physics community. The most recent one was the $W$-boson mass measurement by the CDF Collaboration~\cite{CDF:2022hxs}, which showed a $7\sigma$ departure from the Standard Model (SM) value. Previously, the Fermilab had reported a $4.2\sigma$ deviation in the anomalous magnetic moment of the muon~\cite{Muong-2:2006rrc,*Muong-2:2021ojo} from the SM value computed by the Muon $g-2$ Theory Initiative~\cite{Aoyama:2020ynm,*Aoyama:2012wk,*Aoyama:2019ryr,*Davier:2017zfy,*Keshavarzi:2018mgv,*Colangelo:2018mtw,*Hoferichter:2019mqg,*Davier:2019can,*Keshavarzi:2019abf,*Kurz:2014wya,*Chakraborty:2017tqp,*Borsanyi:2017zdw,*Blum:2018mom,*Giusti:2019xct,*Shintani:2019wai,*FermilabLattice:2019ugu,*Gerardin:2019rua,*Aubin:2019usy,*Giusti:2019hkz,*Melnikov:2003xd,*Masjuan:2017tvw,*Colangelo:2017fiz,*Hoferichter:2018kwz,*Gerardin:2019vio,*Bijnens:2019ghy,*Colangelo:2019uex,*Pauk:2014rta,*Danilkin:2016hnh,*Jegerlehner:2017gek,*Knecht:2018sci,*Eichmann:2019bqf,*Roig:2019reh,*Colangelo:2014qya,*Blum:2019ugy,*Czarnecki:2002nt,*Crivellin:2021rbq}.\footnote{Note that if, instead of the $e^+e^-\to$ hadrons data-driven calculation by the Theory Initiative, the lattice-calculated value of hadronic vacuum polarisation from the BMW Collaboration~\cite{Borsanyi:2020mff}  
is used, the deviation reduces to only  $1.6\sigma$~\cite{DiLuzio:2021uty}.} In the backdrop, there have been the decade-old persistent anomalies in the $R_{K^{(*)}}$ and $R_{D^{(*)}}$ observables of $B$-meson 
decays independently measured by various experimental collaborations~\cite{BaBar:2012obs,*BaBar:2013mob,*LHCb:2014vgu,*LHCb:2017avl,*LHCb:2015gmp,*LHCb:2017smo,*LHCb:2017rln,*Belle:2015qfa,*Belle:2016ure,*Belle:2016dyj,*Belle:2017ilt,*Amhis:2016xyh}.

Various  new-physics explanations of the $W$-boson mass anomaly have already been proposed in the literature~\cite{Fan:2022dck,*Cheung:2022zsb,*Nagao:2022oin,*Kanemura:2022ahw,*Kawamura:2022uft,*Ahn:2022xeq,*Mondal:2022xdy,*Chowdhury:2022moc,*deBlas:2022hdk,*Yang:2022gvz,*Strumia:2022qkt,*Zhang:2022nnh,*Perez:2022uil,*Zheng:2022irz,*Han:2022juu,*Ahn:2022xeq,*Endo:2022kiw,*Heo:2022dey,*Babu:2022pdn,*Gu:2022htv,*DiLuzio:2022xns,*Bahl:2022xzi,*Lee:2022nqz,*Sakurai:2022hwh,*Blennow:2022yfm,*Tang:2022pxh,*Cacciapaglia:2022xih,*Du:2022pbp,*Lu:2022bgw,*Zhu:2022tpr,*Yuan:2022cpw,*Fan:2022yly,*Song:2022xts,*Asadi:2022xiy,*Athron:2022isz,*Heckman:2022the,*Du:2022brr,*Ghoshal:2022vzo,*Crivellin:2022fdf,*Borah:2022obi,*Popov:2022ldh,*Athron:2022qpo}. However, these low-energy anomalies  may not have  independent origins; they could be manifestations of a single high-energy beyond-the-SM theory. It would be fascinating if there were a new-physics scenario which can simultaneously explain the above anomalies and, at the same time, can be tested at the LHC. 

One explanation could be in terms of leptoquarks (LQs). LQs are hypothetical bosons with nonzero lepton and baryon numbers appearing in various new-physics models~\cite{Pati:1974yy,*Schrempp:1984nj,*Georgi:1974sy,*Barbier:2004ez,*Kohda:2012sr}. A recent experiment at the LHCb detector strongly hints towards the existence of LQs~\cite{LHCb:2021trn} (also see Ref.~\cite{Carvunis:2021dss}). Various LQs are known to resolve different anomalies (see, e.g., Refs.~\cite{Bauer:2015knc,*Li:2016vvp,*Hiller:2016kry,*DAmico:2017mtc,*Hiller:2017bzc,*Calibbi:2017qbu,*Becirevic:2018afm,*Angelescu:2018tyl,*Cheung:2020sbq,*Crivellin:2020tsz,*Dorsner:2020aaz,*Aydemir:2022lrq,Aydemir:2019ynb}).  In this paper, we present a simple solution to these anomalies with two TeV-scale scalar leptoquarks (SLQs)---a weak-singlet $S_1\lt(\mathbf{\bar{3}},\mathbf{1},1/3\rt)$ and a weak-triplet $S_3\lt(\mathbf{\bar{3}},\mathbf{3},1/3\rt)$---that agrees with all relevant experimental bounds. The solution is interesting because (a) it is simple and economical as it requires only a few free parameters and (b) it is testable at the LHC; as we shall see, the current LHC data already restrict parts of the parameter space.

The plan of the paper is as follows. In Sec.~\ref{sec:model}, we discuss our $S_1+S_3$ model and the flavour ansatz. We analyse the anomalies in Sec.~\ref{sec:analysis} and present the result of the parameter scan and discuss the relevant experimental bounds in Sec.~\ref{sec:parascan}. Finally, we present our conclusions  in Sec.~\ref{sec:conclu}.

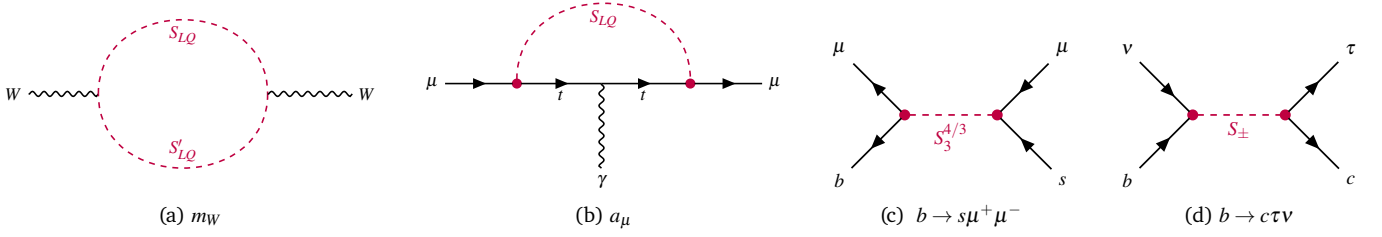
\begin{figure*}
\centering
\subfloat[$m_W$]{
\scalebox{0.75}{\begin{tikzpicture}[thick]
\begin{feynman}
\vertex (i1){\(W\)};
\vertex [right= of i1](a);
\vertex [right= of a](c);
\vertex [right= of c](b);
\vertex [right = of b](f2){\(W\)};
\diagram*{
(i1)--[photon](a),(a)--[scalar,purple,half left,edge label'=\(S_{LQ}\)](b)--[scalar,purple,half left,,edge label'=\(S^\prime_{LQ}\)](a),
(b)--[photon](f2)
};
\end{feynman}
\end{tikzpicture}}
}
\hfill
\subfloat[$a_{\mu}$]{
\scalebox{0.75}{\begin{tikzpicture}[thick]
\begin{feynman}
\vertex (i1){\(\mu\)};
\vertex [right= of i1,purple,dot](a){};
\vertex [right= of a](c);
\vertex [right= of c,purple,dot](b){};
\vertex  [below = of c](r){\(\gamma\)};
\vertex [right = of b](f2){\(\mu\)};
\diagram*{
(i1) -- [fermion](a)--[fermion,edge label'=\(t\)](c)--[fermion,edge label'=\(t\)](b)--[fermion](f2),
(a)--[scalar,purple,half left,edge label'=\(S_{LQ}\)](b),(c)--[boson](r),
};
\end{feynman}
\end{tikzpicture}}
}
\hfill
\subfloat[ $b \to s \mu^+\mu^-$]{
\scalebox{0.82}{\begin{tikzpicture}[thick]
\begin{feynman}
\vertex [purple,dot](c){};
\vertex [above left = of c](a){\(\mu\)};
\vertex [below left = of c](b){\(b\)};
\vertex [right= of c,purple,dot](d){};
\vertex [above right = of d](e){\(\mu\)};
\vertex  [below right = of d](f){\(s\)};
\diagram*{
(c)--[fermion](a), (c)--[fermion](b), (c)--[scalar,purple,edge label'=\(S_{3}^{4/3}\)](d),
(e)--[fermion](d), (f)--[fermion](d),
};
\end{feynman}
\end{tikzpicture}}
\label{fig:bsmumu}}
\hfill
\subfloat[$b \to c \tau \nu $]{
\scalebox{0.82}{\begin{tikzpicture}[thick]
\begin{feynman}
\vertex [purple,dot](c){};
\vertex [above left = of c](a){\(\nu\)};
\vertex [below left = of c](b){\(b\)};
\vertex [right= of c,purple,dot](d){};
\vertex [above right = of d](e){\(\tau\)};
\vertex  [below right = of d](f){\(c\)};

\diagram*{
(c)--[anti fermion](a), (b)--[fermion](c), (c)--[scalar,purple,edge label'=\(S_{\pm}\)](d),
(d)--[fermion](e), (d)--[fermion](f),
};
\end{feynman}
\end{tikzpicture}}
\label{fig:bcnutau}}
\caption{Representative Feynman diagrams contributing to the anomalies. Here, $S_{LQ}$ denotes a scalar LQ.  The vertices with new Yukawa couplings are marked in red.}
\label{fig:Feynman diagram-1}
\end{figure*}

\section{The Leptoquark Model}
\label{sec:model}
\noindent 
Some $S_1+S_3$ models have been studied in different contexts  earlier~\cite{Crivellin:2017zlb,Buttazzo:2017ixm,*Choi:2018stw,*Yan:2019hpm,*Crivellin:2019dwb,*Lee:2021jdr,*Chen:2022hle,*DaRold:2020bib,*Marzocca:2018wcf,*Gherardi:2020det,*Marzocca:2021miv,*Greljo:2021xmg,Saad:2020ihm,Gherardi:2020qhc}. The Yukawa part of the Lagrangian  can be written as~\cite{Buchmuller:1986zs,*Dorsner:2016wpm}
\begin{align}
\mathcal{L}_Y \supset&\ x^{L}_{ij}\bar{Q}_{L}^{C\,i} S_{1} \left(i\sigma_2\right) L_{L}^{j} + x^{R}_{ij}\bar{u}_{R}^{C\,i} S_{1} \ell_{R}^{j}\nn\\&\ + y^{L}_{ij}\bar{Q}_{L}^{C\,i} \left(i\sigma_2\right) \left(\vec\sigma\cdot \vec S_{3}\right) L_{L}^{j} + \textrm{h.c.}, \label{eq:S1S3}   
\end{align}
where we ignore the diquark interactions. The indices $i,j$ denote the generations of the SM fermions, while $Q^{C\,i}_L$ and $L_L^i$ are the $i$th-generation (charge-conjugated) quark and lepton doublets, respectively. The Pauli matrices are denoted by $\sigma_k$'s. In general, the couplings $x^{L}$, $x^{R}$, and $y^{L}$ are $3 \times 3$ complex matrices, but for simplicity, we assume all couplings are real. The triplet $S_3$ has three components with charges $-2/3$, $1/3$, and $4/3$. To produce a shift in $m_W$, we need a mass split among the components. We consider a Higgs portal to mix two different LQs where a mass split appears after the electroweak symmetry breaking (EWSB)~\cite{Dorsner:2019itg}. The relevant terms in the scalar potential are given by
\begin{align}
\mathcal{L}_S \supset&\ - \displaystyle\sum_{i=1,3}\Big[M_{S_i}^2 S_i^\dagger S_i + \lm^H_{S_i} \lt(H^\dagger H\rt)\lt(S_i^\dagger S_i\rt)\nn\\
&\ + \lm_{S_i} \lt(S_i^\dagger S_i\rt)^2 \Big] 
- \Big[\lm H^\dagger\lt(\vec{\sigma}\cdot\vec{S_3}\rt)HS_1^{\ast} + \textrm{h.c.}\Big].
\label{eq:S1S3Higgs}
\end{align}
Here, $H$ is the SM Higgs doublet, and $M_{S_i}^2$, $\lm_{S_i}^H$, $\lm_{S_i}$, and $\lm$ are positive  quantities. The charge-$1/3$ components of the two SLQs ($S_1$ and $S_3^{1/3}$) mix in the presence of a nonzero $\lm$ after EWSB. We get the following mass matrix for the charge-$1/3$ fields:
\begin{equation}
\mc{M}^2 = 
\begin{pmatrix}
M_1^2 & \lm v^2/2  \\
\lm v^2/2 & M_3^2 
\end{pmatrix},
\end{equation}
where $M_i^2 = M_{S_i}^2 + \lm_{S_i}^H v^2$. An orthogonal transformation can diagonalise it:
\begin{equation}
\begin{pmatrix}
S_{-}  \\
S_{+} 
\end{pmatrix} = 
\begin{pmatrix}
\cos\theta & \sin\theta  \\
-\sin\theta & \cos\theta
\end{pmatrix}
\begin{pmatrix}
S_{1}  \\
S_{3}^{1/3}
\end{pmatrix}
\end{equation}
where $S_{\pm}$ are the mass eigenstates with masses $m_{\pm}$, and $\theta$ is the mixing angle:
\begin{align}
M_{\pm}^2 =&\ \dfrac{1}{2}\lt[M_{3}^2 + M_1^2 \pm
\dfrac{1}{2}\sqrt{\lt(M_{3}^2 - M_1^2\rt)^2 + \lt(\lm v^2/2\rt)^2}\rt],\\
\theta  =&\ \frac12\tan^{-1}\left(\frac{\lm v^2/2}{M_1^2 - M_3^2}\right)\in \lt[-\pi/4,\pi/4\rt].
\end{align}

\subsection{Flavour ansatz}
\noindent
We use a flavour ansatz:
\begin{align}
\label{eq:flavansatz}
x^{L/R} =\  
\begin{pmatrix}
0 & 0 & 0 \\
0 & 0 & x^{R}_{23}\\
0 & x^{L/R}_{32} & 0
\end{pmatrix};~~
y^{L} =\  
\begin{pmatrix}
0 & 0 & 0 \\
0 & y^L_{22} & 0\\
0 & y^L_{32} & 0
\end{pmatrix},
\end{align}
which leads us to the following interaction terms:
%
\begin{align}
\label{eq:lagexpli}
\mathcal{L} \supset &
\big[x^{L}_{32}\;\big(V_{td}\ \bar{u}^{C}_{L}\mu_{L} + V_{ts}\ \bar{c}^{C}_{L}\mu_{L} + V_{tb}\ \bar{t}^{C}_{L}\mu_{L}\big)- x^{L}_{32}\;\bar{b}^{C}_{L}{\nu_{\mu}}_L
\nn\\
&\ 
+ x^{R}_{32}\;\bar{t}^{C}_{R}\mu_{R}+x^{R}_{23}\;\bar{c}^{C}_{L}\tau_{L}\big]S_{1} - \big[y^{L}_{32}\;\big(\bar{b}^{C}_{L}{\nu_{\mu}}_L + V_{td}\ \bar{u}^{C}_{L}\mu_{L}  \nn\\ 
&\ + V_{ts}\ \bar{c}^{C}_{L}\mu_{L} + V_{tb}\ \bar{t}^{C}_{L}\mu_{L}\big) + y^{L}_{22}\;\big(\bar{s}^{C}_{L}{\nu_{\mu}}_L + V_{cd}\ \bar{u}^{C}_{L}\mu_{L}  \nn\\
&\ 
+ V_{cs}\ \bar{c}^{C}_{L}\mu_{L} + V_{cb}\ \bar{t}^{C}_{L}\mu_{L}\big) \big]S_3^{1/3} + \sqrt{2}\ y^{L}_{32}\big(V_{td}\ \bar{u}^{C}_{L}{\nu_{\mu}}\nn\\
&\ + V_{ts}\ \bar{c}^{C}_{L}{\nu_{\mu}} + V_{tb}\ \bar{t}^{C}_{L}{\nu_{\mu}}\big)S_3^{-2/3} + \sqrt{2}\ y^{L}_{22}\big(V_{cd}\ \bar{u}^{C}_{L}{\nu_{\mu}} \nn\\
&\ 
+ V_{cs}\ \bar{c}^{C}_{L}{\nu_{\mu}} + V_{cb}\ \bar{t}^{C}_{L}{\nu_{\mu}}\big)S_3^{-2/3} -\sqrt{2}\big(y^{L}_{32}\ \bar{b}^{C}_{L}\mu_{L} \nn\\
&\ 
+ y^{L}_{22}\ \bar{s}^{C}_{L}\mu_{L} \big)S_3^{4/3}.
\end{align}
%
Here, we have assumed the LQ interactions to be aligned with the down quarks and suppressed the neutrino-mixing matrix (since neutrino mixing would not affect the short-ranged experimental measurements of the low-energy observables). Our choice of new couplings is economical, because each plays an essential role in resolving the anomalies simultaneously. (Note that the zeros in the coupling matrices are phenomenological and may not be strictly applicable in specific models~\cite{Bordone:2019uzc,*Bordone:2020lnb}.) The rationales behind the flavour ansatz are as follows:
\begin{enumerate}
\item 
Based on the chiralities of the initial and final muons, the contributions of the SLQs to $\Delta a_\m$ can be split into two parts---chirality-preserving and chirality-flipping. The chirality-preserving terms are small, since they are proportional to $m_\m^2$; whereas the chirality-flipping ones go as $m_\m m_q$ ($q$ is the quark that runs in the loop). Hence, for heavy quarks like the top, the chirality-flipping contribution is bigger than the chirality-preserving one. With $q=t$, the muon $g-2$ discrepancy can be accommodated with perturbative couplings. We primarily consider the $x_{32}^{L/R}$ couplings of the $S_1$ for this purpose. 

\item 
The $S_1$ contributes to the $b\to s\mu\mu$ decay at the loop level, and hence needs large Yukawa couplings to resolve the $R_{K^{(*)}}$ anomalies. Such large couplings would either be ruled out or be in tension with the LHC dilepton data. However, the charge-$4/3$ component of $S_3$, $S_3^{4/3}$ can contribute to $R_{K^{(*)}}$ at the tree level, and it therefore needs relatively smaller couplings. In order to  contribute to the $b\to s\mu\mu$ decay, it needs two nonzero couplings--- namely, $b\mu S_3^{4/3}$ and $s\mu S_3^{4/3}$. We get the required interactions by considering the couplings $y_{32}^L$ and $y_{22}^L$ to be nonzero. (If the $S_3$ couplings are aligned with the up quarks, the two necessary interactions could be generated from only $y_{22}^L$ or $y_{32}^L$. The other essential coupling would then be generated through the CKM mixing. However, due to the small off-diagonal CKM elements, it has to be large, and hence in conflict with the LHC data.)

\item 
We consider the $x_{23}^R$ coupling to generate a positive tree-level contribution to the $R_{D^{(*)}}$ observables from the $S_1$. The coupling allows the charge-$1/3$ LQs to couple with a $\tau$-lepton--$c$-quark pair. To resolve the $R_{D^{(*)}}$ anomalies, the product $x_{32}^Lx_{23}^R$ needs to be order $1$ for TeV-scale LQs. (Note that one could also use the $x_{23}^L$ coupling for this purpose. However, a large $y_{32}^Lx_{23}^L$ or $x_{32}^Lx_{23}^L$ would get into conflict with the current $R_{K}^{\nu\nu}$ measurements unless some additional constraints are enforced~\cite{Crivellin:2017zlb} in this case.)

\end{enumerate}
We show some representative Feynman diagrams contributing to the anomalies in Fig.~\ref{fig:Feynman diagram-1}.

\section{Contribution to the anomalies}\label{sec:analysis}
\subsection{$W$ mass}
\noindent
Due to the $S_1\leftrightarrow S_3^{1/3}$ mixing, a mass split is induced among the components of $S_3$. This shifts the oblique parameters, which induce a shift in the $W$ mass (the corrections to the oblique parameters due to SLQs are studied in Ref.~\cite{Crivellin:2020ukd}). In the $S_1+S_3$ model, the shift of the $T$ parameter, $\Dl T$, is given by
\begin{equation}
\Delta T= \frac{3}{4\pi s_{W}^{2}}\frac{1}{m_{W}^{2}}\lt[F(M_{3},M_{-})c^{2}_{\theta}+F(M_{3},M_{+})s^{2}_{\theta}\rt],
\end{equation}
where $s_W$ is the sine of the Weinberg angle,  $c_\theta=\cos(\theta)$ and $s_\theta=\sin(\theta)$. The function $F(m_a,m_b)$ is given as
\begin{equation}
F(m_{a},m_{b})=m_{a}^{2}+m_{b}^{2}-\frac{2m_{a}^{2}m_{b}^{2}}{m_{a}^{2}-m_{b}^{2}}\text{log}\left(\frac{m_{a}^{2}}{m_{b}^{2}}\right).
\end{equation}
It goes to zero in the limit $m_a=m_b$. The $W$-boson mass is connected to the oblique parameters through the following relation:
\begin{equation}
\hspace{-0.2cm}\Delta m_{W}^{2}=\frac{\alpha_Z c_{W}^{2}m_{Z}^{2}}{c_{W}^{2}-s_{W}^{2}}\lt[-\dfrac{\Dl S}{2} + c_W^2\Delta T + \dfrac{c_{W}^{2}-s_{W}^{2}}{4s_W^2}\Dl U\rt],\hspace{-0.25cm}
\end{equation}
where $\al_Z$ is the fine-structure constant at the $Z$ pole and $c^2_W=1-s_W^2$. The recent CDF measurement puts the $W$-boson mass at $m_W^{\rm CDF}= 80.4335\pm0.0094$ GeV~\cite{CDF:2022hxs}, about $7\sg$ away from its SM value, $m_W^{\rm SM}=80.361 \pm 0.006$ GeV~\cite{ParticleDataGroup:2016lqr}. In our model, the main contribution comes from $\Dl T$; we neglect the smaller shifts in the $S$ and $U$ parameters. The sign of $\Delta M = M_3 - M_1$ is important; we use $\Delta M > 0$ in our analysis. 

\subsection{Muon ${\mathbf g-2}$}
\noindent
The $4.2\sg$ discrepancy in muon $g-2$ translates to $\Dl a_\mu = a_\m^{\rm Exp}-a_\m^{\rm SM} = (2.51\pm0.59)\times 10^{-9}$. In our model, $S_{\pm}$ and $S_3^{4/3}$ contribute to the muon $g-2$. The total contribution to $\Dl a_\mu$ is obtained in Package-X~\cite{Patel:2015tea} as,
\begin{widetext}
\begin{align}
\Delta a_{\mu}=&\ \frac{N_{c}}{16\pi^{2}}\Bigg[-\frac{m_{\mu}m_{t}}{M_{+}^{2}}\left\{\frac{7}{6}+\frac{2}{3}\ln{\left(\frac{m_{t}^{2}}{M_{+}^{2}}\right)}\right\}\Big(2x^{L}_{32}x^{R}_{32}V_{tb}s^{2}_{\theta}+x^{R}_{32}y^{L}_{32}V_{tb}s_{2\theta}+x^{R}_{32}y^{L}_{22}V_{cb}s_{2\theta}\Big) 
+\frac{1}{6}\frac{m_{\mu}^{2}}{M_{+}^{2}}\Big\{\lt(|x^{L}_{32}|^{2}V_t+|x^{R}_{32}|^{2}\rt)s^{2}_{\theta}\nn\\
&\ \quad\quad+\lt(|y^{L}_{32}|^{2}V_t+|y^{L}_{22}|^{2}V_c\rt)c^{2}_{\theta}+x^{L}_{32}y^{L}_{32}V_t s_{2\theta}\Big\} 
-\frac{m_{\mu}m_{t}}{M_{-}^{2}}\left\{\frac{7}{6}+\frac{2}{3}\ln{\left(\frac{m_{t}^{2}}{M_{-}^{2}}\right)}\right\}\Big(2x^{L}_{32}x^{R}_{32}V_{tb}c^{2}_{\theta}-x^{R}_{32}y^{L}_{32}V_{tb}s_{2\theta}-x^{R}_{32}y^{L}_{22}V_{cb}s_{2\theta}\Big)\nonumber\\
&\ \quad\quad+\frac{1}{6}\frac{m_{\mu}^{2}}{M_{-}^{2}}\lt\{\lt(|x^{L}_{32}|^{2}V_t+|x^{R}_{32}|^{2}\rt)c^{2}_{\theta}+\lt(|y^{L}_{32}|^{2}V_t+|y^{L}_{22}|^{2}V_c\rt)s^{2}_{\theta}-x^{L}_{32}y^{L}_{32}V_t s_{2\theta}\rt\}
-\frac{2}{3}\frac{m_{\mu}^{2}}{M_{3}^{2}}\Big(|y^{L}_{32}|^2+|y^{L}_{22}|^2\Big)\Bigg].
\label{eq:amu}
\end{align}
\end{widetext}
Here, $V_t=|V_{td}|^2+|V_{ts}|^2+|V_{tb}|^2$ and $V_c=|V_{cd}|^2+|V_{cs}|^2+|V_{cb}|^2$.  Since the top-quark mass enhances the chirality-flipping terms, the corresponding couplings need not be large,  and thus would have no conflict with the current LHC data. 

\subsection{$R_{K^{(*)}}$ and $R_{D^{(*)}}$ observables}
\noindent
 Our flavour ansatz implies that $S_3^{4/3}$ generates two Wilson operators, $\mc{O}_9 = \lt(\bar{s}\gm_\al P_L b\rt)\lt(\bar{\mu}\gm^\al \mu\rt)$ and $\mc{O}_{10} = \lt(\bar{s}\gm_\al P_L b\rt)\lt(\bar{\mu}\gm^\al\gm^5 \mu\rt)$, relevant for the $R_{K^{(*)}}$ observables. The corresponding coefficients can be expressed in terms of the model parameters,
\begin{equation}
\mc{C}_9 = - \mc{C}_{10} = \dfrac{\pi v^2}{\al V_{tb}V_{ts}^\ast}\dfrac{y_{32}^{L}y_{22}^{L}}{M_{3}^2}.
\end{equation}
The global fit for $\mc{C}_9 = - \mc{C}_{10}$ stands at $-0.39^{+0.07}_{-0.07}$~\cite{Altmannshofer:2021qrr,*Carvunis:2021jga}. The negative value implies both the Yukawa couplings are either positive or negative, since $V_{ts}$ is also negative. There is a small loop-level contribution to $R_{K^{(*)}}$ from $S_{\pm}$, which we ignore. 

Both the charge-$1/3$ components, $S_{\pm}$, contribute to the $R_{D^{(*)}}$ observables at the tree level. From the interactions in Eq.~\eqref{eq:lagexpli}, the coefficient of the operators $\mc{O}_{S_L} = \lt(\bar{c} P_L b\rt)\lt(\bar{\tau} P_L\nu\rt)$ and $\mc{O}_{T_L} = \lt(\bar{c}\sigma^{\mu\nu} P_L b\rt)\lt(\bar{\tau}\sigma_{\mu\nu}  P_L\nu\rt)$ can be written as
\begin{align}
\mc{C}_{S_L} = -4\rho\mc{C}_{T_L} =&\ \dfrac{-1}{4\sqrt{2}V_{cb}G_{\rm F}}\Big[\dfrac{x_{32}^Lx_{23}^Rc^{2}_{\theta}}{M_{-}^2}+\dfrac{x_{32}^Lx_{23}^R s^{2}_{\theta}}{M_{+}^2}\nn\\
&\ +\dfrac{y_{32}^Lx_{23}^Rc_{\theta}s_{\theta}}{M_{-}^2}-\dfrac{y_{32}^Lx_{23}^Rc_{\theta}s_{\theta}}{M_{+}^2}\Big],
\end{align}
where $\rho=\rho(m_b,M_\pm)$ accounts for the modification due to the running strong coupling~\cite{Cai:2017wry}. Figure~4 of Ref.~\cite{Aydemir:2019ynb} shows $\rho$ for a range of the LQ mass scale.
Note that in our case, a second-generation neutrino  would be produced in the $b\to c\ta \n$ process, whereas the SM would produce a third-generation neutrino. Therefore, the $\mc{O}_{S_L}$ and $\mc{O}_{T_L}$ operators in our model will not interfere with the SM one and contribute to the $R_{D^{(*)}}$ observables as~\cite{Iguro:2018vqb}
\begin{align}
r_D \equiv \frac{R_{D}}{R_{D}^{\mathrm{SM}}}\approx\ 1&+ 1.02\ |\mc{C}_{S_L}|^2 + 0.9\ |\mc{C}_{T_L}|^2 \nn\\
&+ 1.49\ \textrm{Re}\lt[\mc{C}_{S_L}\rt] + 1.14\ \textrm{Re}\lt[\mc{C}_{T_L}\rt], \\
r_{D^{*}} \equiv \frac{R_{D^{*}}}{R_{D}^{\mathrm{SM}}} \approx\ 1& + 0.04\ |\mc{C}_{S_L}|^2 + 16.07\ |\mc{C}_{T_L}|^2 \nn\\
&- 0.11\ \textrm{Re}\lt[\mc{C}_{S_L}\rt]
- 5.12\ \textrm{Re}\lt[\mc{C}_{T_L}\rt].
\end{align}
The current averages of the $R_{D^{(*)}}$ anomalies imply that~\cite{Aydemir:2019ynb}
\begin{align}
r_D=1.137\pm0.101\ \mbox{ and}\   
r_{D^*}=1.144\pm0.057.
\end{align}
The $\mc{O}_{S_L}$ and $\mc{O}_{T_L}$ operators also contribute to the $F_L(D^*)$ and $P_{\tau}(D^*)$ observables as,
\begin{align}
f_L(D^*) \equiv \frac{F_{L}(D^*)}{F_{L}^{\textrm{SM}}(D^*)} &\approx\ \frac{1}{r_{D^{*}}}\Big\{1 + 0.08\  |\mc{C}_{S_L}|^2 + 7.02\  |\mc{C}_{T_L}|^2\nn\\
&\ - 0.24\ \textrm{Re}\lt[\mc{C}_{S_L}\rt]- 4.37\ \textrm{Re}\lt[\mc{C}_{T_L}\rt] \Big\},\\
p_{\tau}(D^*) \equiv \frac{P_{\tau}(D^*)}{P_{\tau}^{\textrm{SM}}(D^*)} &\approx\ \frac{1}{r_{D^{*}}}\Big\{1- 0.07\ |\mc{C}_{S_L}|^2- 1.86\times |\mc{C}_{T_L}|^2\nn\\
&\  + 0.22\ \textrm{Re}\lt[\mc{C}_{S_L}\rt]- 3.37\ \textrm{Re}\lt[\mc{C}_{T_L}\rt] \Big\} .
\end{align}

The couplings that contribute to the $R_{K^{(*)}}$ and $R_{D^{(*)}}$ observables also contribute to the flavour-changing neutral-current process $b\to s\nu\nu$, which is loop-induced and suppressed by the Glashow-Iliopoulos-Maiani mechanism in the SM. In our model, the $R_{K}^{\nu\nu}$ observable receives the following correction~\cite{Cai:2017wry}:
\begin{align}
R_{K}^{\nu\nu} &= 1 + \dfrac{\mc{A}^2}{3V_{tb}^2V_{ts}^2}\Bigg\{\lt( \dfrac{y_{32}^Ly_{22}^Ls^{2}_{\theta}}{M_{-}^2}+\dfrac{y_{32}^Ly_{22}^Lc^{2}_{\theta}}{M_{+}^2} +\dfrac{x_{32}^Ly_{22}^L s_{\theta}c_{\theta}}{M_{-}^2}\rt. \nonumber\\
&  \lt.-\dfrac{x_{32}^Ly_{22}^Ls_{\theta}c_{\theta}}{M_{+}^2}\rt)^2\Bigg\} - \dfrac{2\mc{A}}{3V_{tb}V_{ts}}\lt( \dfrac{y_{32}^Ly_{22}^Ls^{2}_{\theta}}{M_{-}^2}+\dfrac{y_{32}^Ly_{22}^Lc^{2}_{\theta}}{M_{+}^2} \rt.\nonumber\\
&\lt.+\dfrac{x_{32}^Ly_{22}^L s_{\theta}c_{\theta}}{M_{-}^2}-\dfrac{x_{32}^Ly_{22}^Ls_{\theta}c_{\theta}}{M_{+}^2}\rt)\label{eq:RKnunu}
\end{align}
where $\mc{A} = \sqrt{2}\pi^2/(e^2G_F|C_L^{\textrm{SM}}|)$ with $C_L^{\textrm{SM}} \approx - 6.38$~\cite{Cai:2017wry}. Note that 
Ref.~\cite{Crivellin:2017zlb} essentially imposed a relation among the couplings through a discrete symmetry to satisfy the $R_{D^{(*)}}$ and $R_{K}^{\nu\nu}$ measurements simultaneously; our solution to that problem is completely different.

\begin{figure*}
\captionsetup[subfigure]{labelformat=empty}
\subfloat[\quad\quad\quad(a)]{\includegraphics[width=0.225\textwidth]{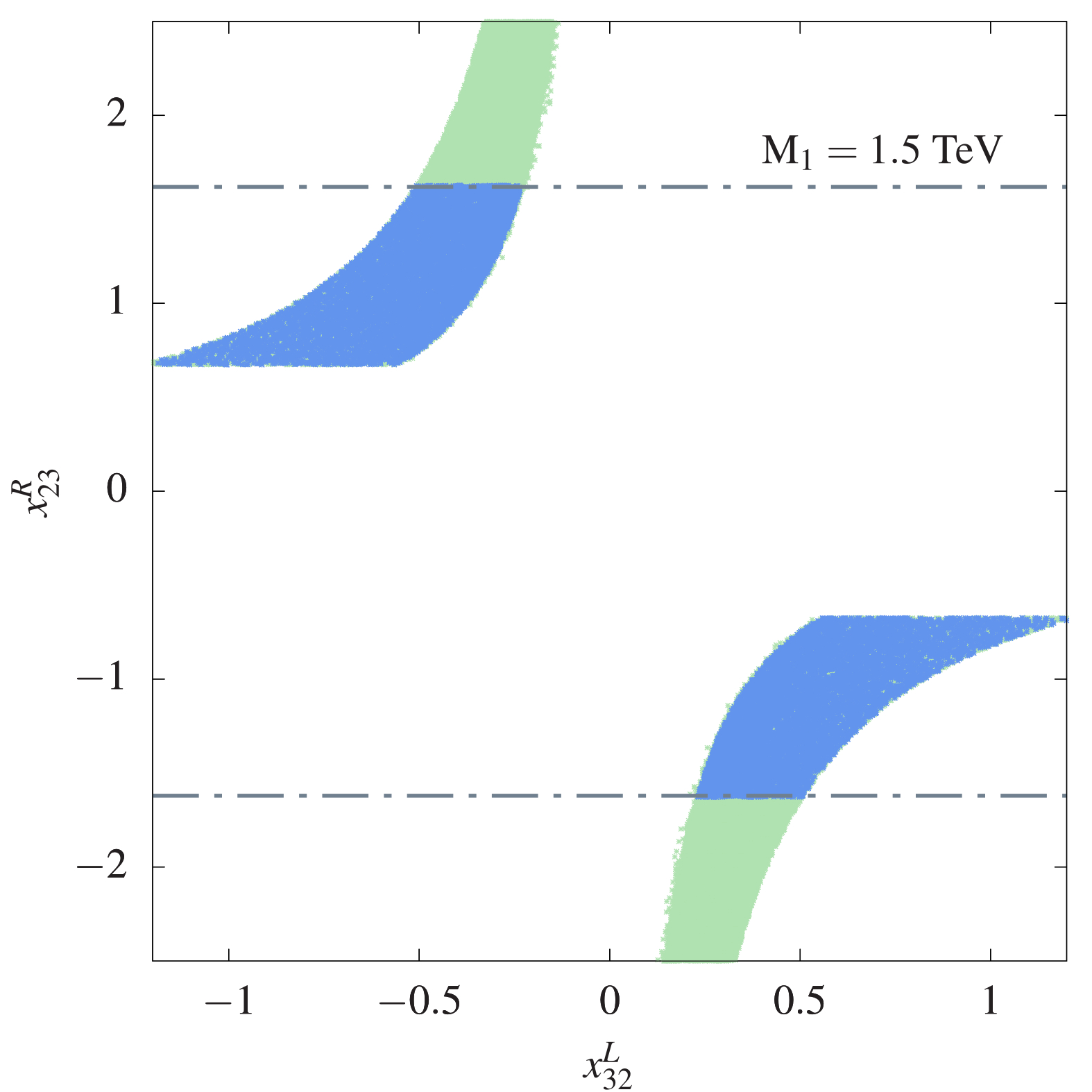}\label{fig:xl32_xr23}}\quad\quad\quad\quad
\subfloat[\quad\quad\quad(b)]{\includegraphics[width=0.225\textwidth]{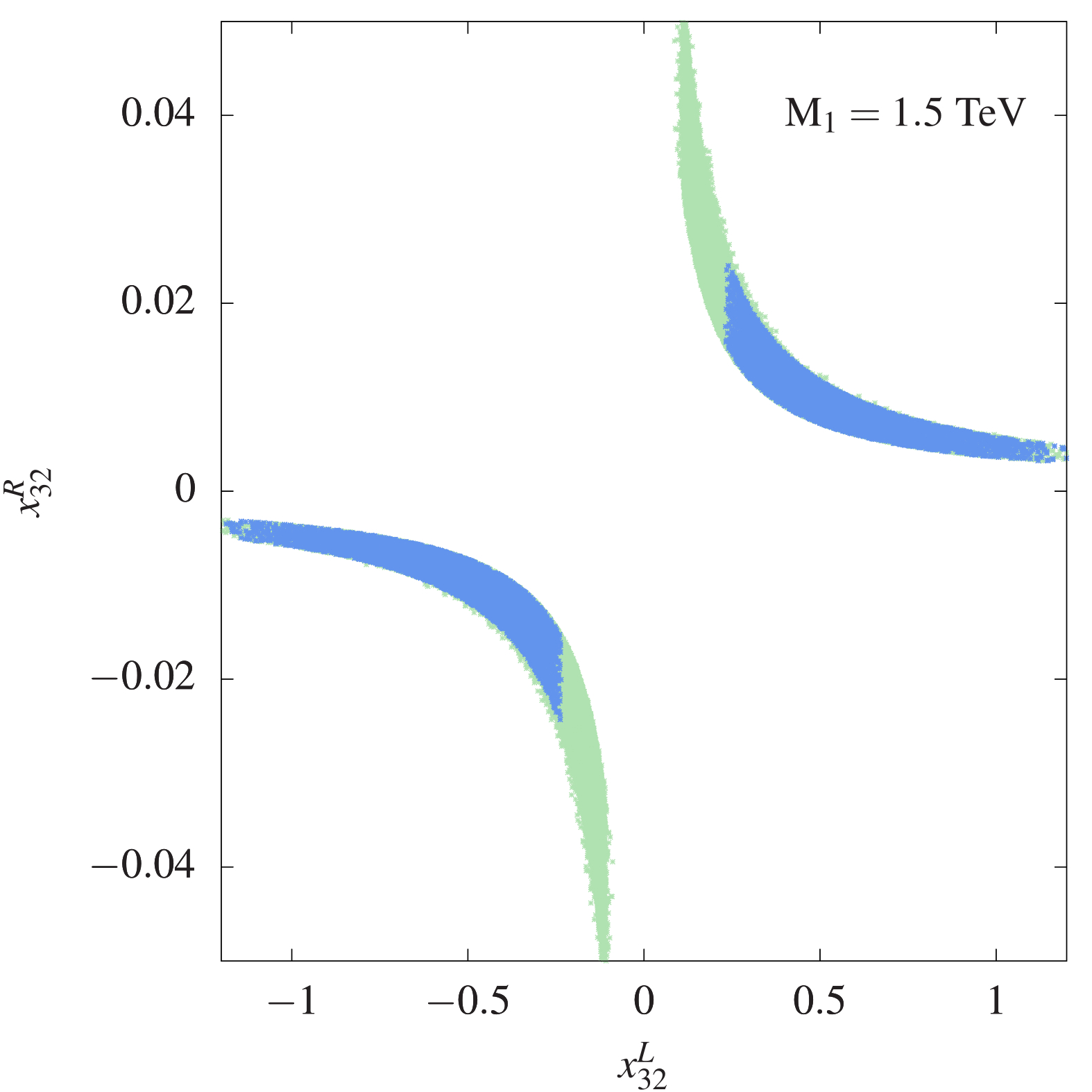}\label{fig:xl32_xr32}}\quad\quad\quad\quad
\subfloat[\quad\quad\quad(c)]{\includegraphics[width=0.225\textwidth]{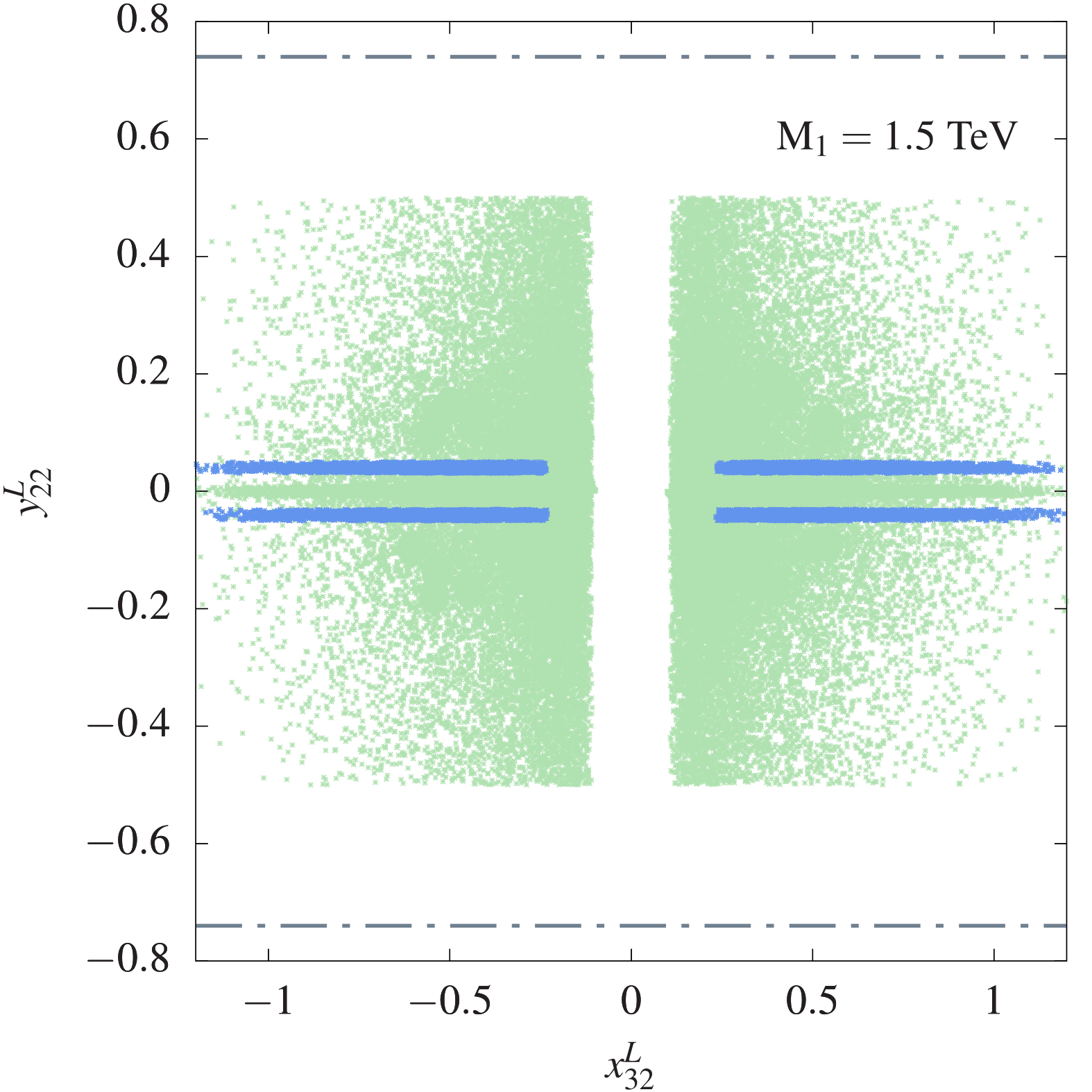}\label{fig:xl32_yl22}}\\
\subfloat[\quad\quad\quad(d)]{\includegraphics[width=0.225\textwidth]{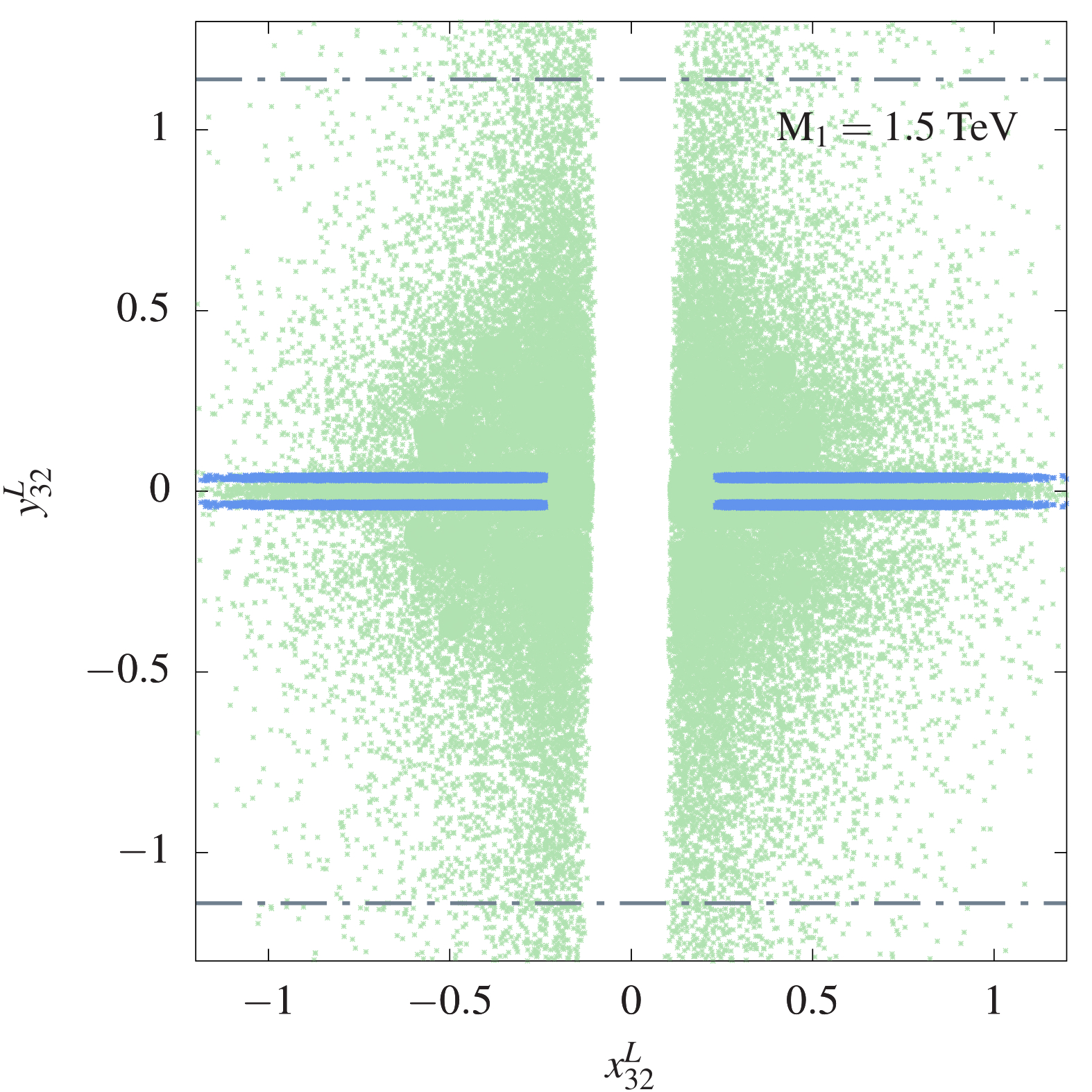}\label{fig:xl32_yl32}}\quad\quad\quad\quad
\subfloat[\quad\quad\quad(e)]{\includegraphics[width=0.225\textwidth]{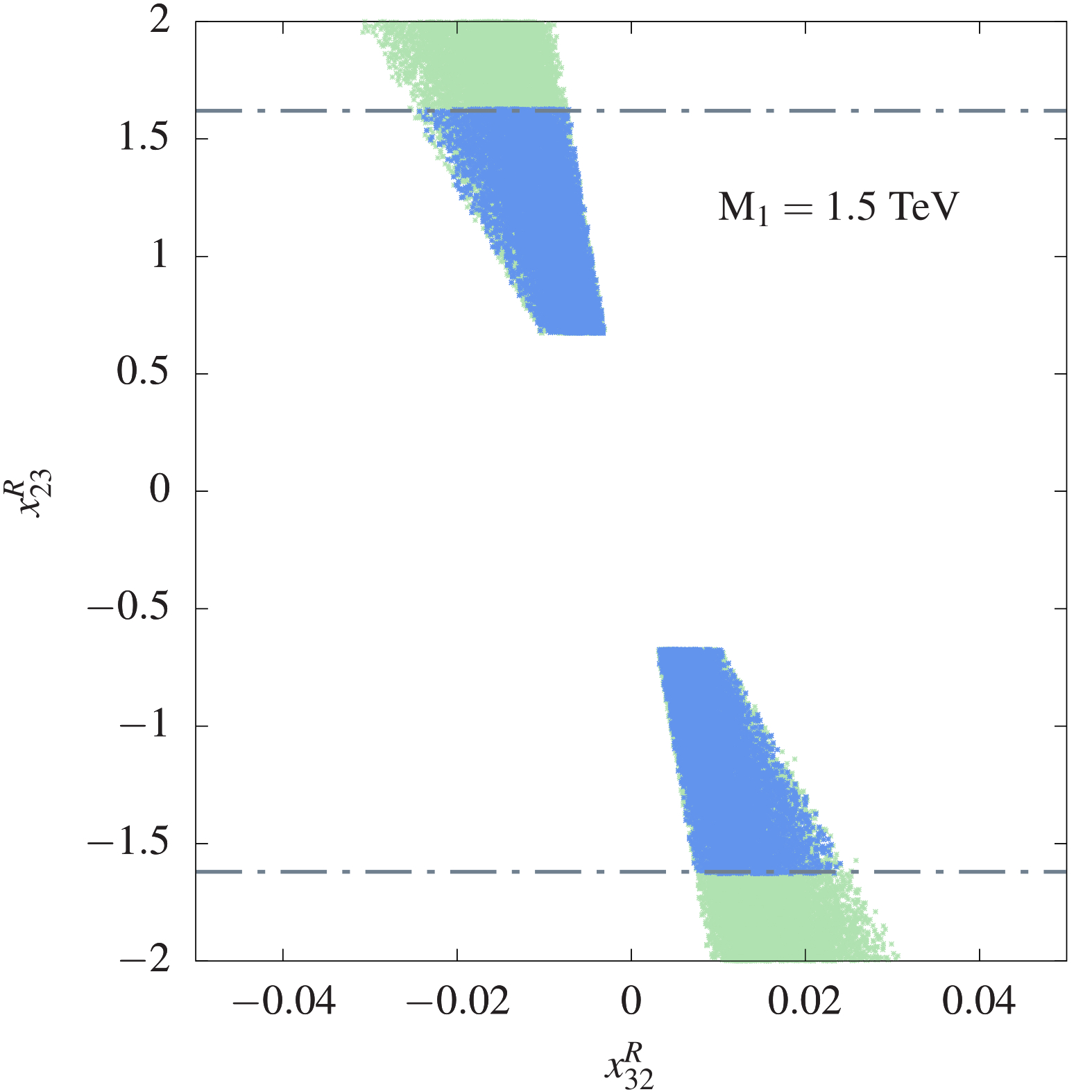}\label{fig:xr32_xr23}}\quad\quad\quad\quad
\subfloat[\quad\quad\quad(f)]{\includegraphics[width=0.225\textwidth]{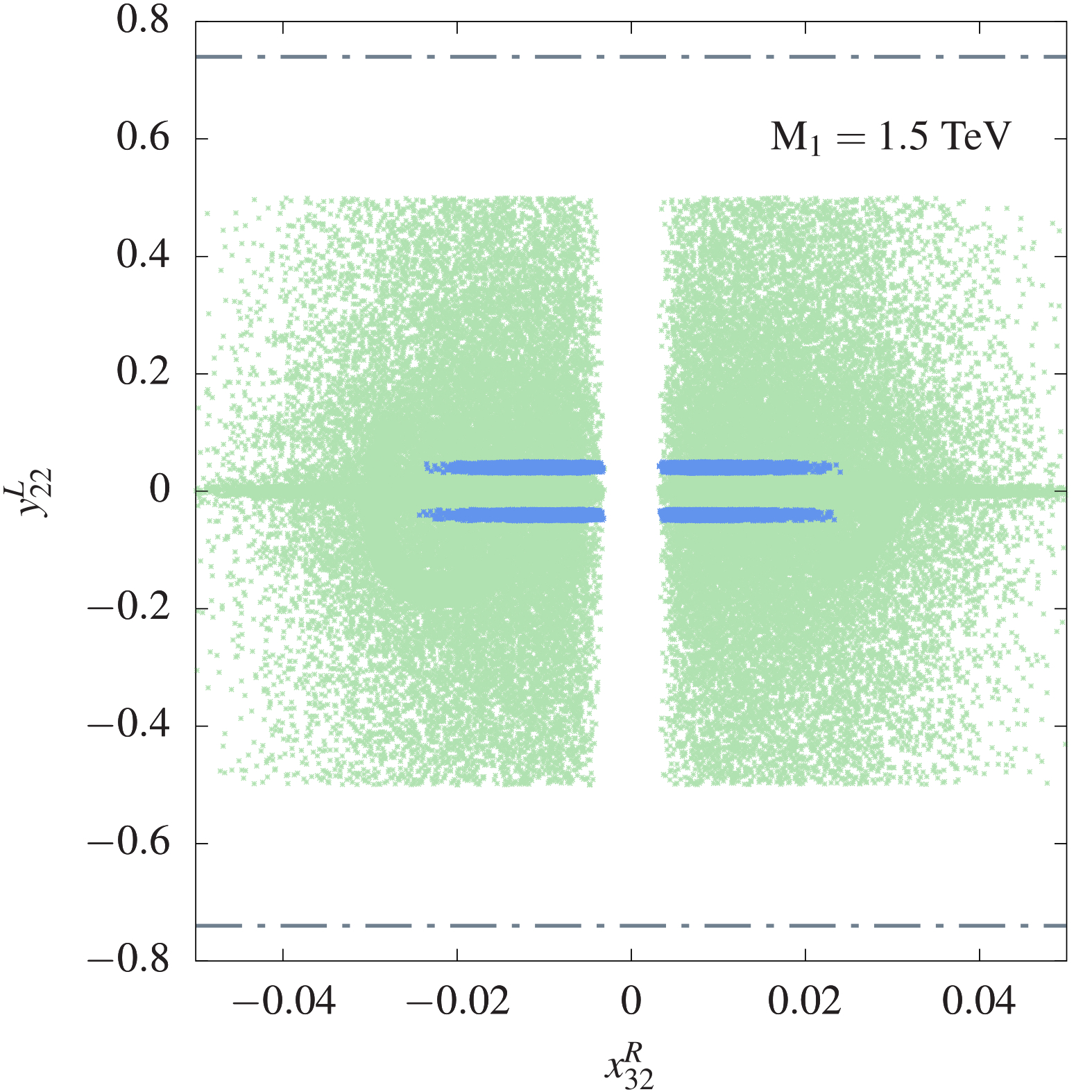}\label{fig:xr32_yl22}}\\
\subfloat[\quad\quad\quad(g)]{\includegraphics[width=0.225\textwidth]{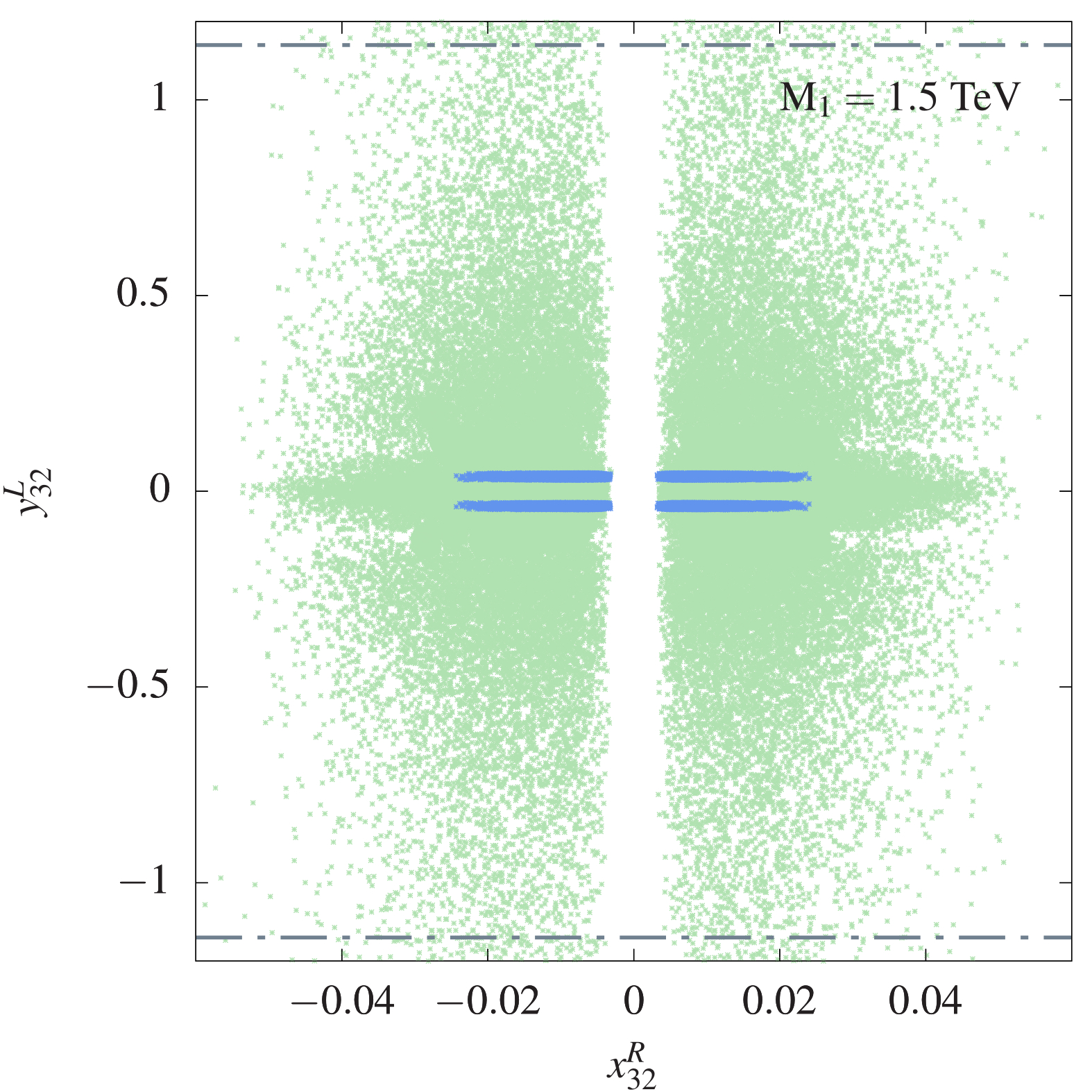}\label{fig:xr32_yl32}}\quad\quad\quad\quad
\subfloat[\quad\quad\quad(h)]{\includegraphics[width=0.225\textwidth]{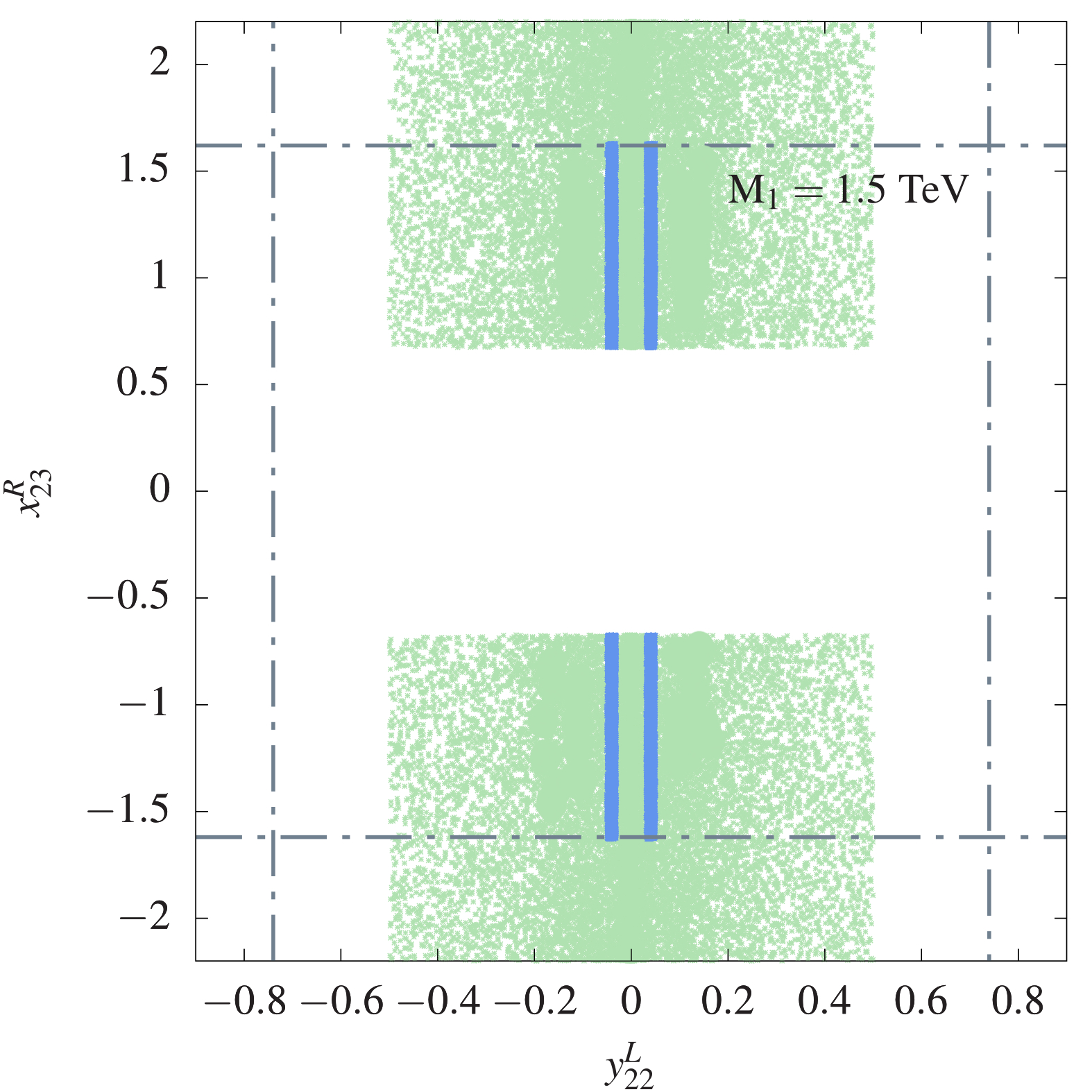}\label{fig:yl22_xr23}}\quad\quad\quad\quad
\subfloat[\quad\quad\quad(i)]{\includegraphics[width=0.225\textwidth]{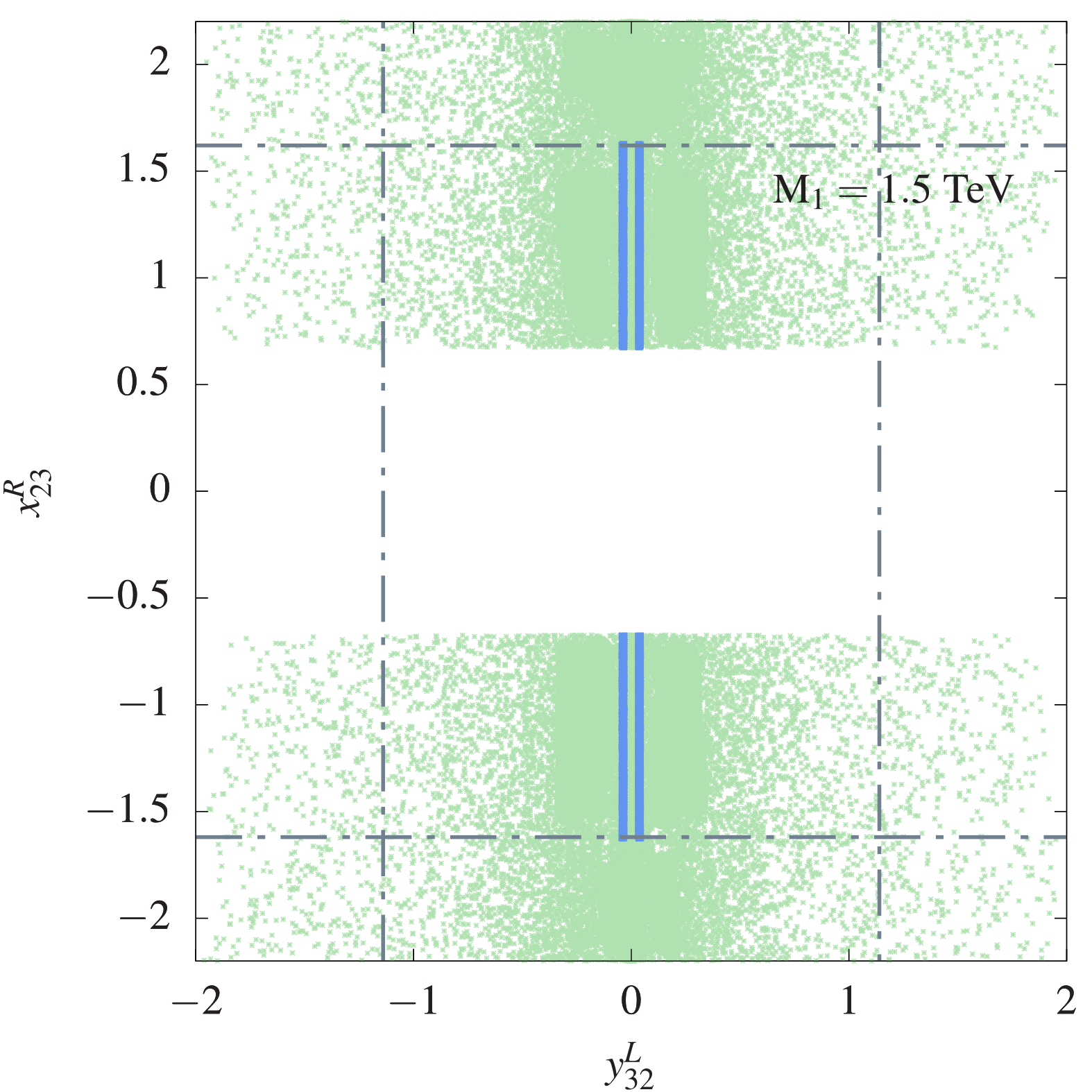}\label{fig:yl32_xr23}}\\
\subfloat[\quad\quad\quad(j)]{\includegraphics[width=0.225\textwidth]{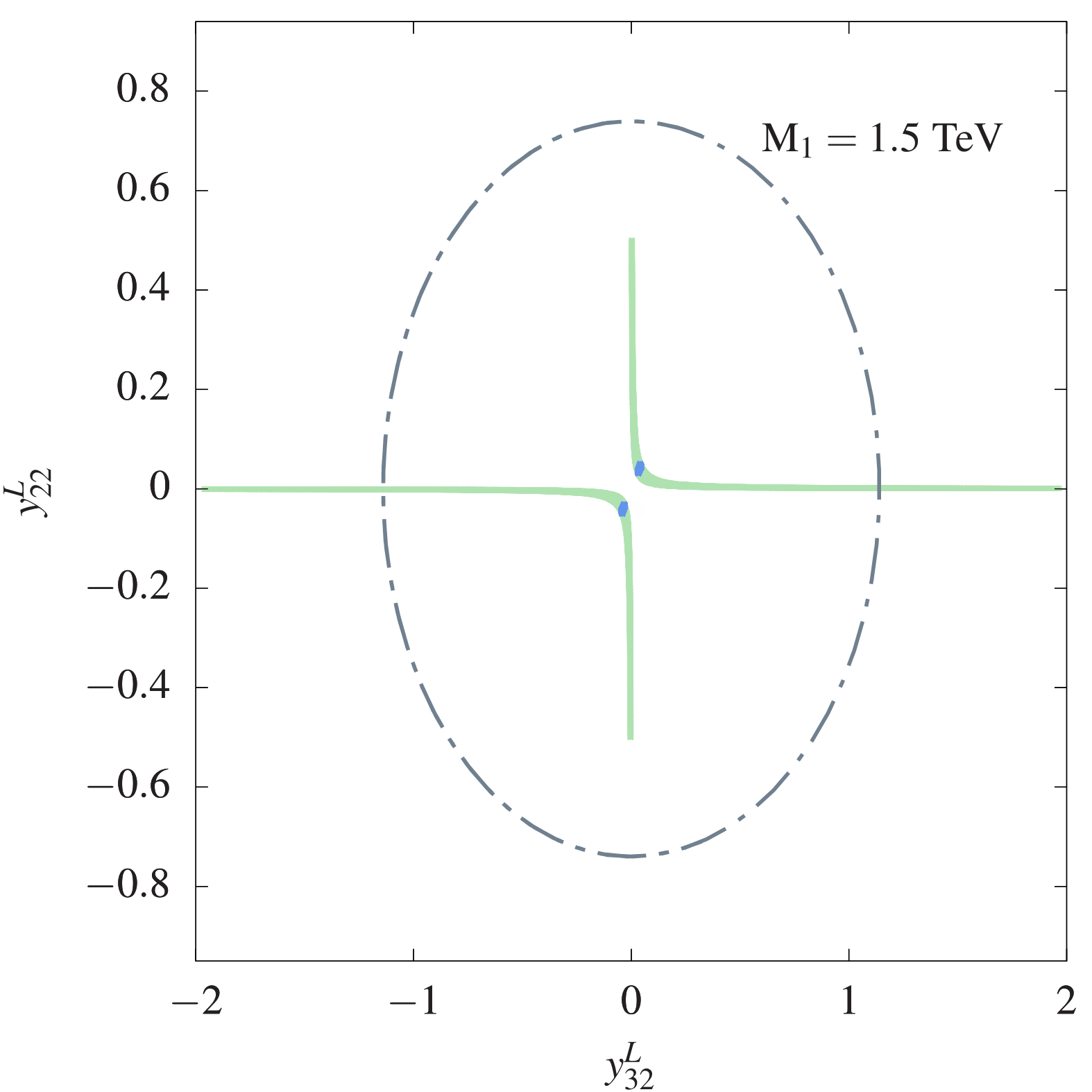}\label{fig:yl32_yl22}}\quad\quad\quad\quad
\subfloat[\quad\quad\quad(k)]{\includegraphics[width=0.225\textwidth]{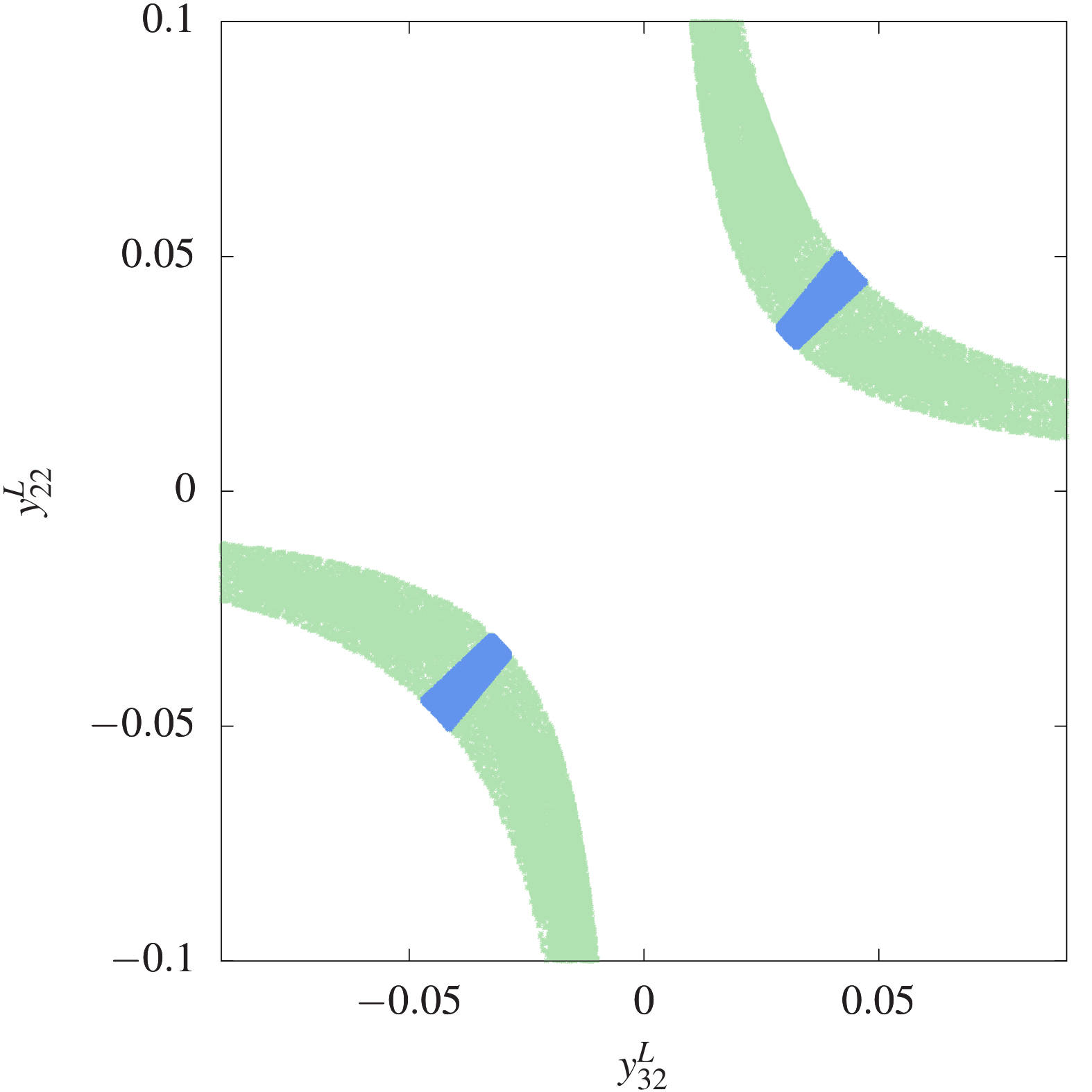}\label{fig:yl32_yl22_magnify}}
\caption{The light-green and blue points belong to the parameter regions where the $1.5$ TeV $S_1+S_3$ model resolves the $W$-mass, muon $g-2$, $R_{K^{(*)}}$, and $R_{D^{(*)}}$ anomalies  without violating the limits shown in Table~\ref{tab:otherobs}. However, once the LHC bounds [the upper limits on couplings from the high-$p_T$ dilepton data (dot-dashed lines) and the direct-search exclusion limits] are applied, only the blue regions survive. Panels (a)--(j) are self-explanatory. Panel (k) shows the relevant part of Panel (j) with higher magnification. }
\label{fig:flavourandLHC}
\end{figure*}
\begin{table}[!t]
\vspace{-7px}\caption{Observables sensitive to the new Yukawa couplings. BR($x\to y$) denotes the branching ratio of the $x\to y$ decay. We use the ``$\star$'' symbol to denote the couplings that contribute through off-diagonal CKM terms. }
{\small\centering\renewcommand{\arraystretch}{1.3}
\begin{tabular*}{\columnwidth}{l @{\extracolsep{\fill}} cr}
\hline
Observable   & Relevant couplings &  Experimental bounds\\
\hline\hline
$R_{K}^{\nu\nu}$ & $x^{L}_{32},y^{L}_{32},y^{L}_{22}$~\cite{Cai:2017wry}& $<2.7$~\cite{Belle:2017oht}\\
BR$(\tau \rightarrow \mu \gamma)$ & $x^{R}_{23}, x^{L\star}_{32},y^{L\star}_{32},y^{L}_{22}$~\cite{Saad:2020ihm} & $<4.4 \times 10^{-8}$\cite{BaBar:2009hkt}\\
BR($B_{c}\rightarrow \tau \bar{\nu})$ & $x^{L}_{32}, x^{R}_{23}, y^{L}_{32}$~\cite{Cai:2017wry}& $\leq 10\%$~\cite{Akeroyd:2017mhr} \\
$f_{L}(D^{*})$  &$x^{L}_{32}, x^{R}_{23},y^{L}_{32}$~\cite{Iguro:2018vqb}& 1.277$\pm$ 0.193~\cite{Bordone:2019vic,Belle:2019ewo}\\
$p_{\tau}(D^{*})$  &$x^{L}_{32}, x^{R}_{23},y^{L}_{32}$~\cite{Iguro:2018vqb}& 0.766$\pm$ 1.093~\cite{Aydemir:2019ynb}\\
$\delta g^{Z}_{\mu_{L}}(Z \rightarrow\mu\mu)$ &$x^{L}_{32}, y^{L}_{32}, y^{L\star}_{22}$~\cite{Cai:2017wry}&$(0.3\pm1.1)\times10^{-3}$~\cite{ALEPH:2005ab} \\
$\delta g^{Z}_{\mu_{R}}(Z \rightarrow\mu\mu)$ &$x^{R}_{32}$~\cite{Cai:2017wry}&$(0.2\pm1.3)\times10^{-3}$~\cite{ALEPH:2005ab}\\
$\delta g^{Z}_{\tau_{R}}(Z \rightarrow\tau\tau)$ &$x^{R}_{23}$~\cite{Cai:2017wry}&$(0.66\pm0.66)\times10^{-3}$~\cite{ALEPH:2005ab}\\
BR$(D^{0}\rightarrow \mu \mu)$ &$x^{L\star}_{32}, y^{L\star}_{32}, y^{L\star}_{22}$\cite{Cai:2017wry}&$<7.6\times 10^{-9}$~\cite{LHCb:2013jyo}\\
BR$(\tau \rightarrow \mu \mu \mu)$ & $x^{R}_{23},x^{L\star}_{32},y^{L\star}_{32},y^{L}_{22}$~\cite{Saad:2020ihm}&$<2.1\times10^{-8}$~\cite{Hayasaka:2010np}\\
BR$(K \rightarrow \mu \nu)$&$x^{L\star}_{32}, y^{L\star}_{32}, y^{L}_{22}$~\cite{Cai:2017wry}&$(63.56\pm0.11)\%$~\cite{Workman:2022ynf}\\
BR$(D_{s} \rightarrow \mu \nu)$&$x^{L\star}_{32}, y^{L\star}_{32}, y^{L}_{22}$~\cite{Cai:2017wry}&$(0.543\pm0.015)\%$~\cite{Workman:2022ynf}\\
BR$(B \rightarrow \mu \nu)$ &$x^{L\star}_{32}, y^{L\star}_{32}, y^{L\star}_{22}$~\cite{Cai:2017wry} & $8.6\times 10^{-7}$~\cite{Workman:2022ynf}\\
$\Delta m(B^{0}_{s}-\bar{B}^{0}_{s})$  &$y^{L}_{32}, y^{L}_{22}$~\cite{Crivellin:2019dwb,Saad:2020ihm}& $\displaystyle (0.993\pm0.158) \Delta m^{\rm SM}_{B_{s}}$~\cite{UTfit:2006onp,ParticleDataGroup:2018ovx} \\
\hline
\end{tabular*}
\label{tab:otherobs}}
\end{table}

\section{Parameter scan and experimental bounds}\label{sec:parascan}
\noindent
We prefer a solution that is directly testable at the LHC. Hence, we consider the LQ mass scale as close to a TeV as possible. As we shall argue below, the lightest LQ the current LHC data allows in our setup is about $1.5$ TeV. We perform a random scan of the five nonzero couplings in Eq.~\eqref{eq:flavansatz} to locate the parameter regions where our $S_1+S_3$ model explains the anomalies simultaneously, keeping the LQ masses fixed at $M_1 = 1.5$ TeV and $M_3 = 1.525$ TeV. The slight mass difference, $\Dl M = M_3 - M_1 \approx 25$ GeV, is necessary to explain the $W$-mass anomaly. Currently, there is no bound on the Higgs-portal coupling, $\lm$, that controls the mixing between $S_1$ and $S_3$---we set it to $1$. 

These five couplings would contribute to  observables other than those mentioned above as well. We list these observables and the experimental bounds on them in Table~\ref{tab:otherobs}. We show the results of the scan with two-dimensional projections in Fig.~\ref{fig:flavourandLHC}. The light-green and blue patches are the regions where the $S_1+S_3$ model resolves the $W$-mass, muon $g-2$, $R_{K^{(*)}}$, and $R_{D^{(*)}}$ anomalies simultaneously while satisfying the bounds in Table~\ref{tab:otherobs} within $2\sg$.

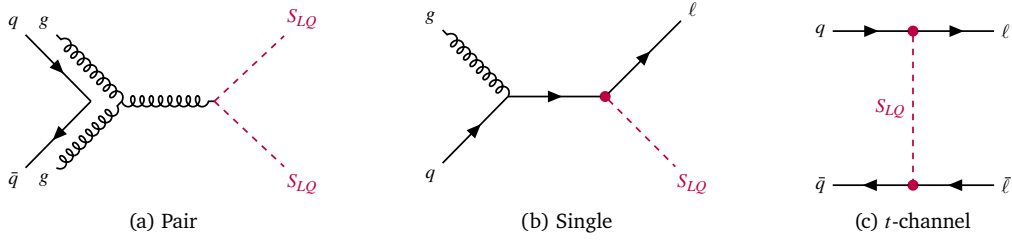
\begin{figure*}
\centering
\subfloat[Pair]{
\scalebox{0.82}{\begin{tikzpicture}[thick]
\begin{feynman}
\vertex (c);
\vertex [above left = of c](a){\(g\)};
\vertex [below left = of c](b){\(g\)};
\vertex [right= of c](d);
\vertex [above right = of d](e){\(\textcolor{purple}{S_{LQ}}\)};
\vertex  [below right = of d](f){\(\textcolor{purple}{S_{LQ}}\)};
\vertex [left = 0.5cm of a](al){\(q\)}; 
\vertex [left = 0.5cm of b](bl){\(\bar q\)}; 
\vertex [left= 0.5cm of c](cl);
\diagram*{
(al)--[fermion](cl), (cl)--[fermion](bl),
(c)--[gluon](a), (b)--[gluon](c), (c)--[gluon](d),
(d)--[scalar,purple](e), (d)--[scalar,purple](f),
};
\end{feynman}
\end{tikzpicture}}
\label{fig:pair}}
\hspace{1cm}
\subfloat[Single]{
\scalebox{0.82}{\begin{tikzpicture}[thick]
\begin{feynman}
\vertex (c);
\vertex [above left = of c](a){\(g\)};
\vertex [below left = of c](b){\(q\)};
\vertex [right= of c,purple,dot](d){} ;
\vertex [above right = 2cm of d](e){\(\ell\)};
\vertex  [below right = 2cm of d](f){\(\textcolor{purple}{S_{LQ}}\)};
\diagram*{
(c)--[gluon](a), (b)--[fermion](c)--[fermion](d)--[fermion](e), (d)--[scalar,purple](f),
};
\end{feynman}
\end{tikzpicture}}
\label{fig:single}}
\hspace{1cm}
\subfloat[$t$-channel]{
\scalebox{0.82}{\begin{tikzpicture}[thick]
\begin{feynman}
\vertex [purple,dot](c){};
\vertex [below= 2.5cm of c,purple,dot](d){} ;
\vertex [left =  of c](a){\(q\)};
\vertex [left = of d](b){\(\bar q\)};
\vertex [right = of c](e){\(\ell\)};
\vertex [right = of d](f){\(\bar \ell\)};
\diagram*{
(a)--[fermion](c), (c)--[fermion](e), (c)--[scalar,purple,edge label'=\(S_{LQ}\)](d),
(d)--[fermion](b), (f)--[fermion](d),
};
\end{feynman}
\end{tikzpicture}}
\label{fig:tchannel}}
\caption{Representative Feynman diagrams for various LQ production processes at the LHC. 
}
\label{fig:production}
\end{figure*}

\subsection{LHC bounds}\label{subsec:lhc}
\noindent
The LHC is insensitive to the small off-diagonal CKM-suppressed couplings. There are direct and indirect bounds on other LQ parameters from the LHC searches~\cite{CMS:2018qqq, CMS:2018oaj,ATLAS:2019qpq,ATLAS:2020dsk, Aad:2020zxo, Sirunyan:2021khd}.  \medskip

\noindent{\bf Direct bounds:} We consider the exclusion bounds from the direct LQ searches on differently charged components separately.

\begin{enumerate}
    \item [(a)] For our flavour ansatz in Eq.~\eqref{eq:flavansatz}, the charge-$1/3$ SLQs would decay to the $t\mu, b\nu, c\tau, s\nu$, and $c\mu$ final states. Now, since $x_{32}^R$ tends to be small in the favoured parameter regions (the light-green regions in Fig.~\ref{fig:flavourandLHC}), individually none of these decay modes can have roughly more than a $50\%$ branching ratio (BR). (This is because if we ignore the CKM-suppressed decays of a charge-$1/3$ LQ, each left-type coupling opens up two decay modes whereas a right-type coupling leads to only one.) Of these, the $c\tau$ mode does not have any direct search limit and, among the others, the strongest one stands at about $1.4$ TeV for a $50$\% BR in the $c\m$ mode~\cite{ATLAS:2020dsk}.

    \item [(b)] The limits on $S^{-2/3}_3$ decaying mainly to $t\nu, c\nu$ final states are weaker---less than a TeV. Hence, our $1.5$ TeV solution trivially agrees with all direct LHC bounds on the charge-$1/3$ and charge-$2/3$ components. 
    
    \item [(c)] The strongest collider bound on our setup comes from the direct limits on $S_3^{4/3}$. In our model, $S_3^{4/3}$ decays to $b\m$ and $s\mu$ final states via $y^L_{32}$ and $y^L_{22}$, respectively. The ATLAS Collaboration has put the lower limit on SLQs that decay to the $b\m$ ($\m j$) state with a $100\%$ BR at $1721$ ($1733$) GeV~\cite{ATLAS:2020dsk}. For the limit in the $b\m$ ($s\m$) mode  to come down to $1.5$ TeV, BR$(S_3^{4/3}\to b(s)\m)\lesssim 0.53~(0.6)$, which forces $|y^L_{32}|$ to be close to $|y^L_{22}|$, since BR$(S_3^{4/3}\to b\m)$~+~BR$(S_3^{4/3}\to s\m)$ $=1$ in our case [see Figs.~\ref{fig:yl32_yl22} and \ref{fig:yl32_yl22_magnify}]. 

\end{enumerate}
 
\noindent{\bf Indirect bounds:} 
The indirect upper bounds on the Yukawa couplings come from the dilepton ($\m\m,~\ta\ta$) data~\cite{Aad:2020zxo, Sirunyan:2021khd}. Since these searches are agnostic about the number of associated jets, the LQs can contribute to the dilepton signals in a number of ways~\cite{Mandal:2018kau, Bhaskar:2021pml} (see Fig.~\ref{fig:production}): 
\begin{enumerate}
    \item [(a)]  They could be pair produced [Fig.~\ref{fig:pair}], $pp \rightarrow S_{LQ}S_{LQ} \rightarrow \ell\ell + 2j$. (Here we use the same notation for a particle and its antiparticle for simplicity; $\ell$ represents a charged light lepton and $j$ denotes a light jet.) The pair production processes are mostly strong-coupling mediated, and hence their contributions to the dilepton signal depend on the new couplings mainly through the BRs. 

    \item [(b)] A LQ can also be produced singly [Fig.~\ref{fig:single}],  $pp \rightarrow S_{LQ} \ell / S_{LQ} \ell j \rightarrow \ell\ell +$ jet(s). The single-production cross sections are proportional to $\kp^2$, where $\kp$ represents the $q\ell S_{LQ}$ coupling involved. (To avoid double counting while calculating the two-body and three-body single-production cross sections, we follow the method explained in Ref.~\cite{Mandal:2015vfa,*Mandal:2012rx}.) 

    \item [(c)] The strongest sensitivity to $\kp$ comes from the $q\bar q^\prime\to \ell\bar\ell$ processes mediated by $t$-channel LQ exchanges [Fig.~\ref{fig:tchannel}], which also interfere with the SM $q\bar q\to Z\to \ell\bar\ell$ processes. For $1.5$ TeV SLQs, the resonant productions are more phase-space suppressed than the nonresonant processes. Hence, at this mass, essentially all of the contribution to the dilepton signals comes from the $\mc O(\kp^4)$ terms from the $t$-channel processes and the $\mc O(\kp^2)$ interference terms. (This can also be seen from Ref.~\cite{Mandal:2018kau} which obtains the limits on $S_1$ couplings.) Hence, we consider only the nonresonant contributions to obtain the indirect limits on the new couplings.
\end{enumerate}

We encode the Lagrangian in Eq.~\eqref{eq:lagexpli} in {\sc FeynRules}~\cite{Alloul:2013bka} to generate the {\sc Universal FeynRules Output}~\cite{Degrande:2011ua} files. We use {\sc MadGraph5}~\cite{Alwall:2014hca} to generate parton-level events from the nonresonant processes. These events are then passed through {\sc Pythia6}~\cite{Sjostrand:2006za} for showering and hadronisation. We use {\sc Delphes v3}~\cite{deFavereau:2013fsa} to simulate the detector environments.

We use two observed distributions from Refs.~\cite{Aad:2020zxo, Sirunyan:2021khd}---the invariant mass distribution of muon pairs and the total transverse mass distribution of tau pairs---to put bounds on the LQ parameters. We apply the same cuts on the SLQ signal events and make a $\chi^2$ estimation of the $2\sg$ limits on the new couplings:
 \begin{align}
        \chi^{2} =& \sum_{i}^{\rm bins}\left(\frac{\mc N_{\rm T}^i(M_{S_{LQ}},\{\kp\})-\mc N_{\rm D}^i}{\Delta \mc N^i}\right)^2
\end{align} 
with
\begin{align}        
        \mc N_{\rm T}^i (M_{S_{LQ}},\{\lm\}) =& \mc N^{nr} (M_{S_{LQ}},\{\kp\}) + \mc N_{\rm SM}^i,\\
        \Delta \mc N^i =& \sqrt{\left( \Delta \mc N^i_{stat}\right)^2 + \left( \Delta \mc N^i_{syst} \right)^2}.\label{eq:errors}
\end{align}
Here, $\mc N^{nr} (M_{S_{LQ}},\{\kp\})$ and $\mc N_{\rm SM}^i$ are the total number of events from the nonresonant production modes passing the selection cuts and the total SM background in the $i$th bin; $\Delta \mc N^i_{stat} = \sqrt{\mc N_D^i}$, and we assume a uniform $10$\% systematic error---i.e., $\Delta \mc N^i_{syst} = \delta^i \times \mc N_D^i$ with $\delta^i = 0.1$. Instead of $\kp$, we use $\{\kp\}$ to indicate that multiple large new couplings can simultaneously contribute to a specific dilepton signal---for example, both the two couplings shown in Fig.~\ref{fig:yl32_yl22} can contribute to the $\m\m$ final state leading to the elliptic upper bound. A more elaborate explanation of the method for obtaining the limits on a number of Yukawa couplings is found in our previous papers~\cite{Mandal:2018kau, Bhaskar:2021pml}.

We show the indirect bounds with dot-dashed lines in Fig.~\ref{fig:flavourandLHC} and mark the common parts between the regions favoured by the anomalies and those agreeing with the LHC limits with blue. From a comparison of the light-green and blue regions in Fig.~\ref{fig:flavourandLHC}, it is clear that our solution is testable at the LHC---the current LHC bounds severely restrict the parameter space of the $1.5$ TeV LQs. The LHC data do not allow the LQ mass scale to be smaller than $\sim1.5$ TeV. This can be seen from Figs.~\ref{fig:yl32_yl22} and \ref{fig:yl32_yl22_magnify} as follows. The $\mc C_9=-\mc C_{10}$ global fit implies that both $|y^L_{32}|$ and $|y^L_{22}|$ cannot be large simultaneously; whereas the LHC direct limits on $S^{4/3}_3$ require $|y^L_{32}|\sim|y^L_{22}|$, severely constricting the parameter space. For a lighter LQ mass, the upper limit on BR$(S_3^{4/3}\to b(s)\m)$ will be tighter and impossible to satisfy. 

 The $1.5$ TeV lower bound points to a tradeoff between the LHC direct bounds and the degree of freedom in the model. By introducing more large couplings, one can reduce the BR of the restrictive modes and reduce the direct limit on the mass scale. However, with more new (large) couplings, the solutions become less appealing or fine-tuned. Note that for LQ masses $1.5$ TeV and above, the indirect limits cannot be diluted significantly by introducing additional coupling(s) as they are essentially independent of the LQ BRs.

\section{Conclusions}
\label{sec:conclu}
\noindent
This paper presents a simple solution to the recent $W$-boson mass anomaly along with the long-standing muon $g-2$, $R_{K^{(*)}}$, and $R_{D^{(*)}}$ anomalies  with a model of $S_1$ and $S_3$ SLQs. While various $S_1+S_3$ models have been seen to address different combinations of these anomalies~\cite{Crivellin:2017zlb,Buttazzo:2017ixm,*Choi:2018stw,*Crivellin:2019dwb,*Yan:2019hpm,*Lee:2021jdr,*Chen:2022hle,*DaRold:2020bib,*Marzocca:2018wcf,*Gherardi:2020det,*Marzocca:2021miv,Saad:2020ihm,Gherardi:2020qhc}, ours is a new solution that is economical (i.e., it has fewer degrees of freedom than the known solutions; see, e.g., Ref.~\cite{Gherardi:2020qhc}), does not require any fine-tuned parameters, and is testable at the LHC.\footnote{Our conclusions remain unchanged if, instead of the $4.2\sigma$, we consider the $a_\m$ anomaly to be $1.6\sigma$ by taking the lattice result~\cite{Borsanyi:2020mff,DiLuzio:2021uty}. In that case, the $|x^R_{32}|\gtrsim 0.03$ regions, which were anyway excluded by the corresponding indirect bound from the LHC data, are disfavoured.} We have shown that our solution satisfies all known low-energy bounds. We also obtained the bounds from the LHC data including those from the direct LQ searches and the high-$p_T$ $\m\m$ and $\ta\ta$ tails.

The testability of the $1.5$ TeV setup at the LHC is clear from the scan shown in Fig.~\ref{fig:flavourandLHC}. If the LQs are heavier, more parameter space will open up---heavier particles are less restricted by the low-energy and LHC bounds. However, heavier SLQs would need larger couplings to accommodate the anomalies. If we limit all the couplings to be perturbative, we get an upper bound on the LQ mass scale---the heaviest it can go (assuming no additional new couplings are turned on) is about $8$ TeV. Of course, if the LQs are significantly heavier than $2$ TeV, the direct searches would not be useful, but the model can still be probed via indirect searches. Future data with better statistics are expected to give better indirect bounds and reduce the upper limit on the LQ mass scale. The large cross-generation couplings (e.g., $x_{23}^R$, $x_{32}^L$) indicate that the model will have exotic signatures (like $c\ta$, $t\m$) at the LHC, and the single productions mediated by these couplings could have good prospects (see, e.g., Refs.~\cite{Mandal:2015vfa,*Chandak:2019iwj,*Bhaskar:2020gkk,*Bhaskar:2021gsy}). They could also be tested indirectly from the high-$p_T$ dilepton or monolepton+$\slashed E_T$ data~\cite{Mandal:2018kau}.

 To summarise, we have presented a testable economical new-physics model that can explain all the anomalies simultaneously without any fine-tuned parameters. Or equivalently, one could also argue that the anomalies together strongly indicate the existence of the LQs. 

\acknowledgements
\noindent
We thank C. Neeraj for a helpful discussion and acknowledge the support from the high-performance computing facility at IISER Thiruvananthapuram, India.

\relax 
\bibliography{leptoquark}
\bibliographystyle{apsrev4-1}

\end{document}